\documentclass[prl,twocolumn]{revtex4}
\usepackage{graphicx}

\begin{document}

\title{Freezing and pressure-driven flow of solid helium in Vycor}

\author{James Day}
\author{Tobias Herman}
\author{John Beamish}

\affiliation{Department of Physics, University of Alberta,
Edmonton, Alberta, Canada, T6G 2J1}

\date{\today}

\begin{abstract}
The recent torsional oscillator results of Kim and Chan suggest a
supersolid phase transition in solid $^4$He confined in Vycor. We
have used a capacitive technique to directly monitor density
changes for helium confined in Vycor at low temperature and have
used a piezoelectrically driven diaphragm to study the
pressure-induced flow of solid helium into the Vycor pores. Our
measurements showed no indication of a mass redistribution in the
Vycor that could mimic supersolid decoupling and put an upper
limit of about 0.003 $\mu$m/s on any pressure-induced supersolid
flow in the pores of Vycor.
\end{abstract}

\maketitle

In a recent torsional oscillator experiment\cite{Kim04-225}, Kim
and Chan observed an unexpected decoupling of $^4$He from a porous
Vycor matrix in a temperature and pressure range (below 175 mK and
around 60 bar) where the helium was solid. The authors described
the helium as a \textquotedblleft supersolid" and speculated that
its non-classical rotational inertia (NCRI) might be associated
with a high vacancy concentration in the confined helium. The same
authors have subsequently observed\cite{Kim04-1941} similar
behavior for bulk helium, implying that supersolidity may be an
intrinsic property of helium. The observation of NCRI in the bulk
solid does not diminish the importance of the Vycor results. Mass
can be transported in bulk crystals via the motion of extended
defects like dislocations or grain boundaries. Such defects may be
essential for supersolidity\cite{Prokofev05-155302,Leggett04-1921}
but would be pinned in small pores and so would not explain the
observed NCRI in Vycor. It is important that other experiments be
done on this system, for example to see whether solid helium
exhibits any of the other unusual flow properties typically
associated with superfluidity, and to rule out alternative
explanations of the NCRI in Vycor. These might include a
persistent liquid layer\cite{Khairallah05a} or a redistribution of
mass due to some other transition in the confined helium. In this
paper we report on experiments in which we used a capacitive
technique to study freezing of helium in Vycor and a new method to
make the first measurements of the flow of solid helium in the
pores in response to external pressure changes.

Vycor is a silica glass with about 28$\%$ of its volume consisting
of a random interconnected network of pores with typical diameter
about 7 nm.  When helium is confined in its pores, a number of
measurements\cite{Beamish83-425, Adams84-2249, Adams87-85} have
shown that the freezing curve is shifted upward by about 10 bar.
The measurements of Adams {\it et al.}\cite{Adams84-2249,
Adams87-85} showed a reduced latent heat of freezing and they
inferred a density change substantially smaller than in bulk. If
this reflects incomplete freezing in the pores, then the
decoupling seen in the torsional oscillator could be occurring in
a liquid layer, rather than in the solid helium. It is also
important to rule out explanations based on a redistribution of
mass. Structural transitions have been seen in a number of
adsorbates in Vycor, including delayering in an argon layer near
the pore surface\cite{Wallacher01-104202} and crystallographic
transitions in oxygen and argon\cite{Molz93-5741, Brown98-1019}.
Also, solid argon and krypton have been seen\cite{Silva02-155701}
to migrate out of the pores well below their freezing
temperatures. Such effects can change the oscillator's moment of
inertia and mimic superfluid decoupling, as was shown for
hydrogen\cite{Schindler96-11451, Dertinger97-r14689} where a
dewetting transition expelled mass from the Vycor.  In the
experiments described in this paper, we used a capacitive
technique to study the density changes associated with freezing of
helium in Vycor and at the lower temperatures where Kim and Chan
observed supersolidity. Our measurements confirm that the density
change associated with freezing is smaller than in bulk helium but
show that it is independent of pressure. This implies that, if
there is a remaining liquid layer, then it must be very difficult
to freeze. We saw no evidence that solid helium spontaneously
entered or left the pores at low temperatures, ruling out mass
redistributions due to non-superfluid transitions as the
explanation of the torsional oscillator results. It is therefore
very interesting to see how solid helium flows in response to a
pressure gradient. By suddenly increasing the pressure in a cell
containing a Vycor sample, we were able to monitor the
pressure-induced flow of solid helium in the pores. Near the
melting temperature, solid helium did flow into the pores but the
rate decreased rapidly with temperature; below about 700 mK no
flow was detected.  Our experiments extended below 50 mK, well
into the temperature range where Kim and Chan observed NCRI. If
the helium in Vycor is a supersolid, then either it does not
respond to pressure differences or superflow occurs at a rate far
slower than the critical velocities of the torsional oscillator
measurements.

Our Vycor sample was a disc with a diameter of 12.7 mm and a
thickness d = 0.52 mm, onto which we evaporated circular copper
electrodes (100 nm thick, 0.71 cm$^2$ area) to form a capacitor.
Before depositing the electrodes, we dusted the Vycor with 40
micron cobalt powder (held in place by a magnet behind the
sample). After deposition, the powder was removed, leaving an
electrically continuous electrode with perforations (about 10$\%$
of the area) to allow the helium easy access to the pores. At 4.4
K the empty sample had a capacitance C$_V$ = 3.7257 pF, roughly
what would be expected from the manufacturer's quoted dielectric
constant for Vycor (3.1 at room temperature). If helium admitted
to the pores acted as a uniform dielectric, then the capacitance
change, C$_V$ would be proportional to the Vycor's porosity,
$\phi$ , and to $\epsilon$$_{He}$-1, the helium's contribution to
the dielectric constant within the pores. Since
$\epsilon$$_{He}$-1  is proportional to the helium density,
capacitance changes would provide a direct measure of the amount
of helium in the sample. In a real porous medium, the contribution
of a pore fluid to the dielectric constant depends on pore
geometry through depolarization effects\cite{Pelster99-9214}, but
measurements with Ar and CO in Vycor have
shown\cite{Wallacher02-014203} that, except for very thin adsorbed
films, this can be accounted for by including a simple geometric
parameter so that the capacitance change is still proportional to
the change in adsorbate density. We have confirmed this for liquid
helium via a 1.8 K adsorption isotherm. For fillings greater than
about two monolayers, the Vycor capacitance increased linearly
with the amount of helium adsorbed.

For our initial freezing measurements, the Vycor capacitor was
sealed into a copper pressure cell which included an {\it in situ}
Straty-Adams pressure gauge. The cell had a volume much larger
than the Vycor pores, so the bulk helium acted as a reservoir
which kept the pressure essentially constant when the helium in
the pores froze. Crystals were grown using the blocked capillary,
constant volume technique. Temperatures were measured with a
calibrated germanium thermometer above about 50 mK, with $^{60}$Co
nuclear orientation and $^3$He melting curve thermometers for
calibration at lower temperatures.  The pressure and helium
density were measured capacitively using an automatic bridge
operating at 1 kHz (Andeen-Hagerling 2550A). Most measurements
were made with a 15 V excitation, after confirming that the few nW
of dissipation in the Vycor capacitor (its resistance was greater
than 3x10$^{10}$ ohms) did not cause any measurable sample heating
down to 30 mK.

\begin{figure}
\includegraphics[width=\linewidth]{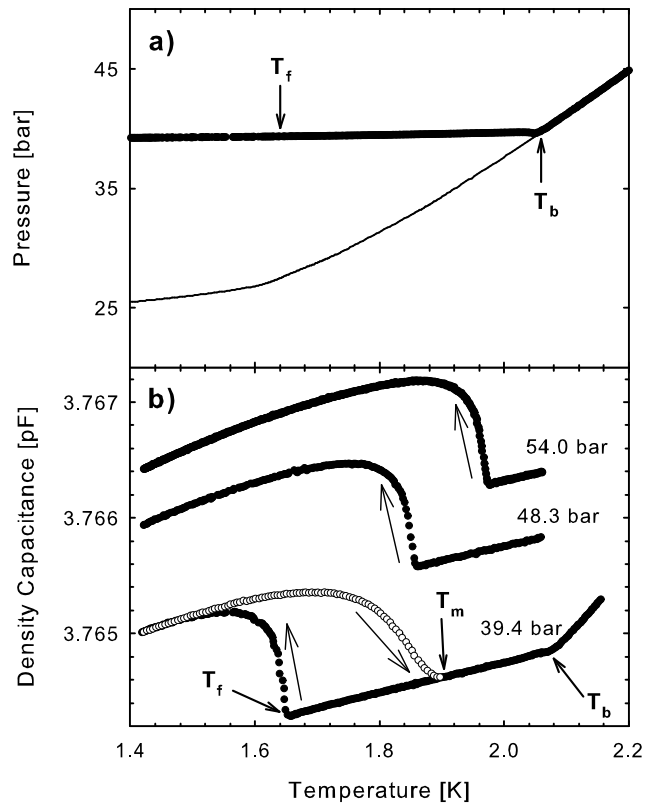}
\caption{Freezing and melting of helium in Vycor.  (a)
Thermodynamic path (pressure vs. temperature) during cooling.
System follows the $^4$He melting curve (solid line) until bulk
freezing is complete at T$_B$ = 2.05 K, P = 39.4 bar.  (b) Lower
curve: capacitance during cooling (solid symbols) and warming
(open symbols) at 39.4 bar.  Upper curves show the corresponding
cooling data at 48.3 and 54.0 bar. } \label{fig:thermpath}
\end{figure}

Figure~\ref{fig:thermpath}(a) shows the thermodynamic path during
a typical measurement. The bulk helium began to freeze at 2.75 K
and 66.7 bar and the pressure then decreased, following the
melting curve down to the point marked \textquotedblleft T$_B$"
(2.05 K, 39.4 bar) where bulk freezing was complete. At lower
temperatures the pressure remained nearly constant and the helium
in the Vycor pores did not begin to freeze until the point marked
\textquotedblleft T$_F$" (1.64 K). Figure~\ref{fig:thermpath}(b)
shows the corresponding capacitance, C$_V$, which reflects the
helium density in the pores.  Along the melting curve the
capacitance decreased, since liquid helium left the pores as the
pressure in the cell dropped, as can be seen in the lowest (39.4
bar) curve. The slower decrease between T$_B$ and T$_F$ is just
the background temperature dependence of the dielectric constant
of Vycor, due to \textquotedblleft two level systems" (TLS) in the
glass. Freezing in the pores was marked by the sudden increase in
C$_V$ at T$_F$, due to the larger density of solid helium.  When
the sample was later warmed (open symbols) the helium melted at
higher temperature, with melting complete at T$_M$ = 1.86 K. The
suppression of freezing and the hysteresis between freezing and
melting are common features of adsorbates in small pores.

By extrapolating and subtracting the background temperature
dependence of C$_V$ we can extract the jump in capacitance,
$\Delta$C$_V$, associated with freezing in the pores. The jump in
Fig.~\ref{fig:thermpath}(b), $\Delta$C$_V$ = 0.0011 pF, is about
2.8$\%$ of the 0.0395 pF capacitance change due to filling and
pressurizing the sample with liquid helium, less than half the
6$\%$ density increase when bulk helium freezes at this
pressure\cite{Grilly62-250}. This may be due to an
\textquotedblleft inert layer" at the pore walls which does not
participate in freezing and melting or may indicate that some of
the helium remains liquid. In the latter case, we might expect the
fraction of the helium which freezes to increase with pressure,
with a correspondingly larger capacitance change. We made
measurements at pressures ranging from 31.7 bar (where no freezing
was seen down to 30 mK) up to 54 bar.
Figure~\ref{fig:thermpath}(b) includes freezing data at several
pressures, all showing the same capacitance change, 0.0011 pF.  If
a persistent liquid layer remains after helium in the pores
freezes, then it is remarkably stable and insensitive to pressure.

\begin{figure}
\includegraphics[width=\linewidth]{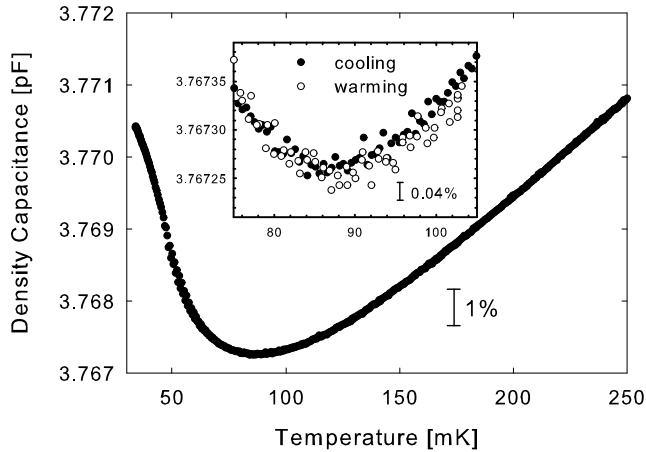}
\caption{Low temperature capacitance of the Vycor/solid helium
sample.  Bar shows the capacitance change for a 1$\%$ change in
the density of the helium in the pores.  Insert is a blow up of
the data around the capacitance minimum at 88 mK and includes data
taken during warming from 30 mK (open symbols).} \label{fig:lowT}
\end{figure}

We next cooled the Vycor sample containing solid helium (at 39.4
bar) to look for any change in helium density that might mimic
superfluidity in a torsional oscillator. Figure~\ref{fig:lowT}
shows the capacitance data at low temperatures. The smooth mimimum
at 88 mK is typical of dielectric glasses and reflects coupling to
the TLS, not changes in the helium density. For example, we saw
the same behavior when the pores contained liquid helium at
saturated vapor pressure. If there was a low temperature
transition which resulted in helium being expelled from the pores,
then it would show up as a sudden decrease in capacitance, but we
saw no such change in the range below 200 mK where Kim and Chan
saw decoupling. The bar in Fig.~\ref{fig:lowT} shows the change
that would be expected if 1$\%$ of the helium were to leave the
pores.  Since movement of helium in or out of the pores might
occur slowly (Kim and Chan observed time constants of order 1 hour
for their oscillator period to stabilize), we waited overnight at
our lowest temperature (about 30 mK) and then warmed our sample.
The insert in Fig.~\ref{fig:lowT} compares the capacitance during
warming (open symbols) to the initial data during cooling. Within
the resolution of our measurements, there is no difference,
demonstrating that the density of the solid helium is constant to
within 0.04$\%$ at low temperatures. Motion of solid helium into
or out of the Vycor pores cannot explain Kim and Chan's
observations, strengthening their interpretation in terms of
supersolidity.

Since our measurements rule out some of the most obvious
alternative explanations of the decoupling observed for solid
helium in Vycor, it becomes interesting to see whether solid
helium exhibits any of the other unusual flow properties of a
superfluid. In our second experiment, by suddenly increasing the
pressure in a cell containing the same Vycor sample, we were able
to monitor the pressure induced flow of solid helium in the pores.
Since thermally activated vacancies can transport mass in a
pressure gradient\cite{Beamish91-9314}, we do expect to see flow
at temperatures near the melting point of the helium in the pores,
but this flow rate should decrease rapidly with temperature.

\begin{figure}
\includegraphics[width=\linewidth]{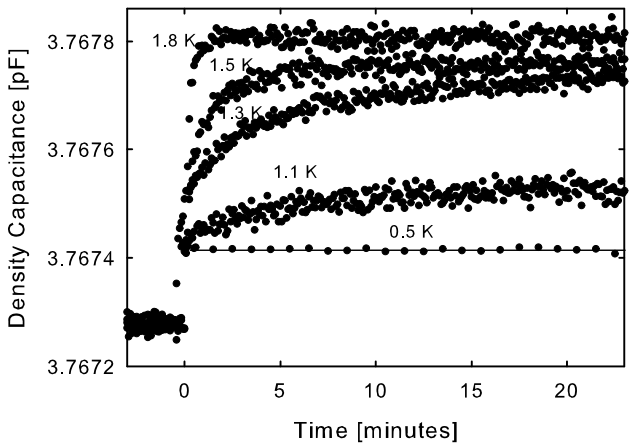}
\caption{Capacitance response to a rapid compression of the
surrounding helium.  From top to bottom, the curves correspond to
temperatures 1.8, 1.5, 1.3, 1.1 and 0.5 K.  Horizontal line
through the 0.5 K data is a guide to the eye.}
\label{fig:squeezes}
\end{figure}

For the pressure/flow measurements we built a beryllium copper
\textquotedblleft squeezing cell" with a flexible diaphragm
machined into one end and an external piezoelectric actuator
designed to compress the helium by up to 1$\%$. We again started
at a pressure high enough to completely freeze the helium in the
pores (at T$_F$ = 2.05 K for the 57 bar data shown below) and
cooled to a temperature between 2 K and 30 mK. We then suddenly
(in about 10 seconds) compressed the helium by applying a voltage
to the piezoelectric actuator while monitoring the helium density
in the Vycor. Figure~\ref{fig:squeezes} shows the results of such
\textquotedblleft squeezes" at five temperatures between 1.8 and
0.5 K. At 1.1 K and above, the capacitance (i.e. the solid helium
density in the pores) responded to the pressure step in two
stages. First, there was an immediate capacitance jump of about
0.133 fF, which occurred within the measurement time of our
capacitance bridge, i.e. during the 10 seconds taken to increase
the pressure. Second, there followed a slower,
temperature-dependent increase. The time constant associated with
the slower increase varied from less than 30 seconds at 1.8 K to
more than an hour at 1.1 K. Below 700 mK (e.g. the 500 mK data in
Fig.~\ref{fig:squeezes}) there was no measurable capacitance
change following the initial jump.

The initial 0.133 fF jumps in Fig.~\ref{fig:squeezes} are simply
due to the elastic compression of the capacitor. Even if no helium
flows into the pores, a pressure change  $\Delta$P will
elastically compress the capacitor and produce a geometric change
$\Delta$C$_V$/C$_V$. This temperature independent change is
immediate and can be calculated from Vycor's dielectric and
elastic constants (Young's modulus E = 1.8x10$^{10}$ Pa; Poisson's
ratio = 0.20).  For uniaxial compression in our cell, we expect
$\Delta$C$_V$/C$_V$ = (1.0x10$^{-10}$ Pa$^{-1}$)$\Delta$P; our
0.133 fF jump corresponds to a pressure increase of about 3.5 bar.
If solid helium subsequently flows into the Vycor to equalize the
pressures after compression, then the capacitance will increase
further, but at a slower rate which depends on the flow velocity.
This capacitance change depends on the compressibility of the
helium in the pores, which can be found from the data of
Fig.~\ref{fig:thermpath}b. Between 48.4 and 54 bar, we estimate
the solid's compressibility (at 1.45 K) as 2.0x10$^{-8}$
Pa$^{-1}$, slightly less than the corresponding value for bulk
helium (2.3x10$^{-8}$ Pa$^{-1}$ at 54 bar\cite{Jarvis68-320}. For
a 3.5 bar pressure step, equilibrating the pressure inside and
outside the pores would produce a change of about 3x10$^{-4}$ pF,
roughly what we observe after the initial jump.

\begin{figure}
\includegraphics[width=\linewidth]{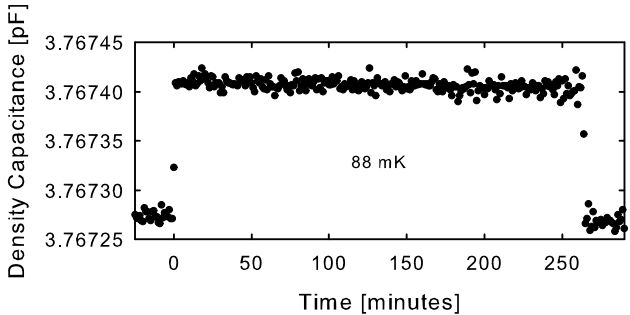}
\caption{Capacitance change for a compression at 88 mK, followed
by a decompression 260 minutes later.} \label{fig:lowTsqueeze}
\end{figure}

The flow-induced capacitance changes in Fig.~\ref{fig:squeezes}
occurred more slowly as the temperature was reduced. This is
consistent with mass transport via a thermally activated process,
presumably the diffusion of vacancies in the solid helium or in a
disordered layer at the pore walls. Above 1.3 K, the capacitances
approached similar final values within the time shown; at 1.1 K
the changes continued for much longer and we did not wait long
enough to determine the asymptotic value. At 0.5 K we saw no flow
at all. The flow behavior depended slightly on the thermal history
of the sample and differences between the final capacitance values
may reflect defect creation associated with deformation of the
bulk solid and annealing at the higher temperatures. Although the
results of Fig.~\ref{fig:squeezes} are not systematic enough to
provide a precise activation energy (the data between 1.1 and 1.8
K indicate a value around 9 K), the essential result is that solid
helium near its melting point flows in Vycor when an external
pressure is applied, but this flow is negligible at temperatures
below about 0.7 K.

The most interesting question is whether the solid helium in the
Vycor responds to a pressure difference when cooled below 175 mK
(the range where Kim and Chan saw decoupling).
Figure~\ref{fig:lowTsqueeze} shows our capacitance results at 88
mK when the pressure was raised, held for about 4 hours, and then
returned to its original value. By taking data at the capacitance
minimum of Fig.~\ref{fig:lowT} (88 mK), we eliminated effects of
the small temperature changes caused by heating in the
piezoelectric actuator. There is no indication of any density
change inside the Vycor following the initial capacitance jump.
About 0.5$\%$ of the helium decoupled in Kim and Chan's Vycor
measurements. If this fraction were to flow from the surface to
the center of our sample at their critical velocity (of order 100
$\mu$m/s), then a 1$\%$ density change outside the pores would be
transmitted throughout the pores within a few seconds.
Figure~\ref{fig:lowTsqueeze} shows that any pressure-induced
helium flow in our experiments must occur at much lower speeds.
Assuming that helium can flow into the Vycor through the
electrode's perforations and at its edges (about 30$\%$ of the
sample's outer surface), we find that the flow velocity must be
less than about 0.003 $\mu$m/s. We extended our squeezing
measurements down to 48 mK with no indication of mass flow. Below
this temperature, dissipation in the piezoelectric actuator heated
the cell slightly and prevented accurate measurements.

The NCRI observed in Kim and Chan's torsional oscillator
measurements appears to be a fundamental property of solid helium
at low temperatures. Our measurements rule out alternative
explanations of their results based on redistribution of mass in
Vycor rather than supersolid decoupling. However, we do not see
any evidence of pressure induced flow in the temperature range
where they observed supersolidity.  This is consistent with
previous experiments by Greywall\cite{Greywall77-1291} which put a
similar limit (0.002 $\mu$m/s, using Kim and Chan's bulk
supersolid fraction\cite{Kim04-1941}, 1.5$\%$) on pressure-induced
flow of bulk solid helium through capillaries. If a supersolid
exists, then its flow properties must be quite different from that
of superfluids, since the chemical potential difference created by
a pressure change does not appear to produce superflow.

This work was supported by the Natural Sciences and Engineering
Research Council of Canada.

\bibliography{pressure}

\end{document}